\documentclass[12pt,twoside]{article}

\usepackage{amsmath,amssymb,slashed}
\usepackage{graphicx,color}
\usepackage{ctable}
\usepackage[sort&compress]{natbib}

\bibpunct{[}{]}{,}{n}{}{,}

\newcommand{\tr}{{\rm tr}} 

\newcommand{\twovect}[2]{\begin{pmatrix} #1 \\ #2 \end{pmatrix}}
\newcommand{\Mpd}{{M_{\slashed{B}}^{\rm min}}}

\title{On the Consistency of Perturbativity and Gauge Coupling Unification}
\author{Joachim Kopp\footnote{Email: jkopp@mpi-hd.mpg.de} ,
        Manfred Lindner\footnote{Email: lindner@mpi-hd.mpg.de} ,
        Viviana Niro\footnote{Email: niro@mpi-hd.mpg.de} ,
        Thomas E.~J.~Underwood\footnote{Email: Thomas.E.J.Underwood@mpi-hd.mpg.de} \\[0.2cm]
         {\it Max--Planck--Institut f\"{u}r Kernphysik}, \\
         {\it Postfach 10 39 80, 69029 Heidelberg, Germany}}
\date{September 16, 2009}

\makeatletter
\def\@maketitle{%
     \renewcommand{\thefootnote}{\alph{footnote}}
     \newpage
     \vspace*{0.5em}
     \begin{center}%
     \let \footnote \thanks
       {\Large\bf \@title \par}%
       \vskip 1.0em%
       {\normalsize
         \lineskip .5em%
         \begin{tabular}[t]{c}%
           \@author
         \end{tabular}\par}%
       \vskip 0.7em%
       {\normalsize \@date}%
     \end{center}%
     \par
     \vskip 0.5em}
\makeatother

\makeatletter
\renewcommand\section{\@startsection {section}{1}{\z@}%
                                   {-3.5ex \@plus -1ex \@minus -.2ex}%
                                   {2.3ex \@plus.2ex}%
                                   {\normalfont\large\bfseries}}
\makeatother

\begin{document}

\maketitle

\begin{abstract}
  We investigate constraints that the requirements of perturbativity and gauge
  coupling unification impose on extensions of the Standard Model and of the
  MSSM. In particular, we discuss the renormalization group running in several
  SUSY left-right symmetric and Pati-Salam models and show how the various
  scales appearing in these models have to be chosen in order to achieve
  unification. We find that unification in the considered models occurs
  typically at scales below $\Mpd = 10^{16}$~GeV, implying potential conflicts
  with the non-observation of proton decay. We emphasize that extending the
  particle content of a model in order to push the GUT scale higher or to
  achieve unification in the first place will very often lead to
  non-perturbative evolution.  We generalize this observation to arbitrary
  extensions of the Standard Model and of the MSSM and show that the
  requirement of perturbativity up to $\Mpd$, if considered a valid guideline
  for model building, severely limits the particle content of any such model,
  especially in the supersymmetric case.  However, we also discuss several
  mechanisms to circumvent perturbativity and proton decay issues, for example
  in certain classes of extra dimensional models.
\end{abstract}

\section{Introduction}

Among the many interesting open problems in particle physics is the question
whether the Standard Model (SM) gauge coupling constants unify at some high
energy scale. Even though such Grand Unification does not occur in the SM, one
of its best motivated extensions, namely the Minimal Supersymmetric Standard
Model (MSSM), does predict Grand Unification at $M_{\rm GUT} \simeq
10^{16}$~GeV~\cite{Amaldi:1991cn}, opening up the possibility to embed the
model into a Grand Unified Theory (GUT) based, for example, on the gauge group
$SO(10)$.

Grand Unification is also possible in more complex models: Amaldi et al.\ have
identified several extensions of the SM or MSSM particle content that would
lead to gauge coupling unification~\cite{Amaldi:1991cn,Amaldi:1991zx}, and
Lindner and Weiser have performed a similar study in left-right symmetric
models~\cite{Lindner:1996tf}. Recently, Calibbi et al.\ have derived a set of
``magic fields'' that can be added to the MSSM without spoiling
unification~\cite{Calibbi:2009cp}. Model-dependent studies have been carried
out for example by Shaban and Stirling~\cite{Shaban:1992vv} for a non-SUSY
left-right symmetric model and by Perez-Lorenzana and Mohapatra for models
with extra dimensions~\cite{PerezLorenzana:1999qb}. A model independent study
of non-SUSY left-right symmetric models has also been carried out by
Perez-Lorenzana et al.~\cite{PerezLorenzana:1998rj}. Recently Aranda et
al. carried out a study of extended Higgs sectors in SUSY GUTs, paying
attention to the constraints coming from the requirements of perturbativity
and freedom from anomalies~\cite{Aranda:2009wh}.

Our aim in this work is to emphasize that the often adopted requirements that
(1) all gauge couplings remain perturbative up to the GUT scale $M_{\rm GUT}$
and (2) $M_{\rm GUT}$ is large enough to suppress proton decay beyond the
experimental limit severely constrain extensions of the SM or the MSSM. Even
though non-perturbativity is not a principle problem but only a practical one,
and proton decay can be suppressed even for low $M_{\rm GUT}$ in suitably
constructed GUTs, we will in this paper accept both (1) and (2) as valid
guidelines for model building, and investigate in detail the constraints that a
model has to fulfil in order to be compatible with them.

In sec.~\ref{sec:RGE}, we will begin by introducing the formalism of RGEs in
order to fix our notation. We will then proceed to a detailed investigation of
Grand Unification in left-right symmetric and Pati-Salam models in
secs.~\ref{sec:LR} and \ref{sec:PS}, respectively.  There, we will discuss how
the various energy scales appearing in these models are constrained by the
requirement of successful perturbative unification compatible with proton decay
bounds. In sec.~\ref{sec:general}, we will generalize our results, and discuss
perturbativity issues in arbitrary extensions of the SM and the MSSM.  Finally,
in sec.~\ref{sec:ways-out} we will outline how perturbativity and proton decay
constraints can be circumvented by more elaborate model building constructs
such as extra dimension. We will summarize our results and conclude in
sec.~\ref{sec:conclusions}.

\section{Renormalization group evolution of gauge coupling constants}
\label{sec:RGE}

The dependence of the gauge coupling constants $g_i$ on the energy scale
$\mu$ for a theory with gauge group $G = \prod_i G_i$ is given at one-loop
order by
\begin{align}
  16 \pi^2 \, \frac{dg_i(t)}{dt} = b_i \, [g_i(t)]^3 \,,
  \label{eq:rge}
\end{align}
where $t = \ln(\mu/\mu_0)$ and $\mu_0$ is an arbitrary renormalization scale.
The coefficients $b_i$, which are determined by the particle content of the
model, have for non-supersymmetric models the form~\cite{Jones:1981we}
\begin{align}
  b_i = \sum_{R} s(R) \, T_i(R) - \frac{11}{3} C_{2i} \,.
  \qquad \text{(non-SUSY models)}
  \label{eq:bi}
\end{align}
Here, the sum runs over all representations of the gauge group factor
$G_i$, counted according to their multiplicity in the model. For example,
in the Standard Model (SM) with its six left-handed quarks and six right-handed
quarks, the 3-dimensional representation of $SU(3)_c$ has a multiplicity of 12.
The Dynkin index $T_i(R)$ of the representation $R$ of $G_i$ is defined
by $\tr[t_i^a(R) \, t_i^b(R)] = \delta^{ab} T_i(R)$, with
$t_i^a(R)$ being the generators of $G_i$ in the representation $R$. If
$G_i = U(1)$, $T_i(R) = [q(R)]^2$, where $q(R)$ is the charge
corresponding to the representation $R$.  The coefficient $s(R)$ has the value
$2/3$ if $R$ is a multiplet of chiral fermions, while $c(R) = 1/3$ if $R$ is a
multiplet of complex scalars.  Finally, $C_{2i}$ is the quadratic Casimir
operator of the adjoint representation of $G_i$.  For supersymmetric
models, eq.~\eqref{eq:bi} has to be replaced by
\begin{align}
  b_i = \sum_{R} T_i(R) - 3 \, C_{2i} \,.
  \qquad \text{(SUSY models)}
  \label{eq:bi-SUSY}
\end{align}
The solution of the one-loop renormalization group equation \eqref{eq:rge} can be
written in the form
\begin{align}
  \alpha_i^{-1}(t) = \alpha_i^{-1}(t_0) - \frac{1}{2\pi} b_i \, (t - t_0) \,,
  \label{eq:alpha}
\end{align}
where $\alpha_i = [g_i(t)]^2 / 4\pi$.

An important observation and one of the central points of this paper is that
adding new (non-singlet) matter particles to a given model will always
\emph{increase} at least one of the $b_i$, and hence will lead to larger
values for the corresponding $\alpha_i(t)$ at $t > t_0$. For sufficiently
large particle content, $\alpha_i(t)$ will reach the non-perturbative
regime at relatively low scales.

All results presented in this paper will be based on the above one-loop RGEs.
In the case of weak couplings this approximation is certainly justified, but
when approaching the non-perturbative regime, $\alpha_i^{-1} \lesssim 1$,
higher order effects will become relevant. Nevertheless, in the context of our
study it is sufficient to define the non-perturbativity scale as the scale at
which the one-loop approximations breaks down or, more specifically, as the
scale at which the one-loop value of at least one of the $\alpha_i^{-1}$
becomes negative.  (Of course, the physical $\alpha_i^{-1}$ will always remain
positive, and it is just the invalidity of the one-loop approximation that can
lead to negative values.)

Also, note that we neglect threshold corrections, which lead to an uncertainty
of typically one order of magnitude for the non-perturbativity scale and the
GUT scale.

\section{Grand Unification in Left-Right Symmetric Models}
\label{sec:LR}

We now illustrate the constraints that perturbativity and unification place on
models with large particle content by considering three different left-right
(LR) symmetric extensions of the Standard Model: (i) A non-supersymmetric
model with Higgs triplets~\cite{Mohapatra:1977mj,Mohapatra:1979ia,Mohapatra:1980yp,
Deshpande:1990ip}, (ii) the ``minimal'' SUSY LR model~\cite{Babu:2008ep}, and
(iii) a slightly extended SUSY LR model~\cite{Aulakh:1997ba,Aulakh:1997fq}.
All three models have in common that quarks and leptons reside in the following
representations under $SU(3)_c \times SU(2)_L \times SU(2)_R \times
U(1)_{B-L}$:
\begin{align}
  Q(3,2,1,\tfrac{1}{3})      &= \twovect{u}{d}          &
  Q^c(3^*,1,2,-\tfrac{1}{3}) &= \twovect{d^c}{-u^c}     \\
  L(1,2,1,-1)                &= \twovect{\nu_e}{e}      &
  L^c(1,1,2,1)               &= \twovect{e}{-\nu_e} \,.
  \label{eq:LR-matter}
\end{align}

\emph{(i) Non-SUSY LR model with triplet Higgs~\cite{Mohapatra:1977mj,
Mohapatra:1979ia,Mohapatra:1980yp,Deshpande:1990ip}.}
In the non-supersymmetric case, the LR symmetry is broken down to the Standard
Model at a scale $M_{\rm LR}$ by Higgs triplets
\begin{align}
  \Delta(1,3,1,2) \qquad\text{and}\qquad \Delta^c(1,1,3,-2) \,.
  \label{eq:LR-triplets-1}
\end{align}
The second of these acquires a vev of order $M_{\rm LR}$ and thus breaks
$SU(2)_R \times U(1)_{B-L}$ down to $U(1)_Y$, while the first one is required
only to keep the particle content left-right symmetric. Fermion masses are
generated by a Higgs bidoublet
\begin{align}
  \Phi(1,2,2,0) \,,
  \label{eq:LR-bidoublet}
\end{align}
with a vev of the order of the electroweak scale. Even though we do not need to
worry about the details of the symmetry breaking mechanism in order to study
the renormalization group evolution of the model, it is crucial to know the
mass scales of all particles.  A detailed investigation~\cite{Deshpande:1990ip}
shows that all Higgs particles in the model have masses of the order of $M_{\rm
LR}$, except for an $SU(2)_L$ doublet emerging from the bidoublet $\Phi$ and
playing the role of the SM Higgs boson.  Note that, even though the model is
generically non-supersymmetric, a high scale supersymmetrization at a scale
$M_{\rm SUSY} > M_{\rm LR}$ is imaginable.

\emph{(ii) Minimal SUSY LR model~\cite{Babu:2008ep}.} The Higgs sector of this
model is given by that of the non-SUSY model (i) (with all fields promoted to
superfields), supplemented by two additional triplets
\begin{align}
  \bar{\Delta}(1,3,1,-2) \qquad\text{and}\qquad \bar{\Delta}^c(1,1,3,2) \,
  \label{eq:LR-triplets-2}
\end{align}
required for anomaly cancellation and a singlet
\begin{align}
  S(1,1,1,0) \,
  \label{eq:LR-singlet}
\end{align}
to ensure charge and $R$-parity conservation. Moreover, in the SUSY case two
Higgs bidoublets $\Phi_1$ and $\Phi_2$ are required to allow for non-vanishing
quark and lepton mixing angles. Of these Higgs superfields, four doublets as
well as the doubly charged components of $\Delta^c$ and $\bar{\Delta}^c$ are
light and have masses of $\mathcal{O}(M_{\rm SUSY})$. In this study, we will
make the simplifying assumption that these fields all have the same mass,
$M_{\rm SUSY}$, whereas in practice, for low values of $M_{\rm SUSY}$ some of
the fields are required to have slightly higher masses to satisfy constraints
on, for example, flavour changing neutral currents. We have checked this
simplification has a very minimal effect on our results.

\emph{(iii) Non-minimal SUSY LR model~\cite{Aulakh:1997ba,Aulakh:1997fq}.}
The particle content of the non-minimal model is similar to that of the minimal
SUSY-LR model, with the singlet $S$ being replaced by two triplets
\begin{align}
  \Omega(1,3,1,0) \qquad\text{and}\qquad \Omega^c(1,1,3,0) \,,
  \label{eq:LR-triplets-3}
\end{align}
uncharged under $U(1)_{B-L}$. Breaking of the left-right symmetry proceeds
in two steps in this model:
\begin{align}
  SU(2)_R \times U(1)_{B-L}
  \xrightarrow{M_{\rm LR}}
  U(1)_R \times U(1)_{B-L}
  \xrightarrow{M_{B-L}}
  U(1)_Y \,.
  \label{eq:LR-breaking-4}
\end{align}
The set of light particles includes the usual MSSM Higgses at the electroweak
scale, the neutral components of $\Delta^c$ and $\bar{\Delta}^c$ at $M_{B-L}$,
and the $\Omega$ field at $\max(M_{B-L}^2 / M_{\rm LR}, M_{\rm
SUSY})$~\cite{Aulakh:1997fq}.

\begin{figure}
  \begin{center}
    \begin{tabular}{cc}
      \includegraphics[width=8cm]{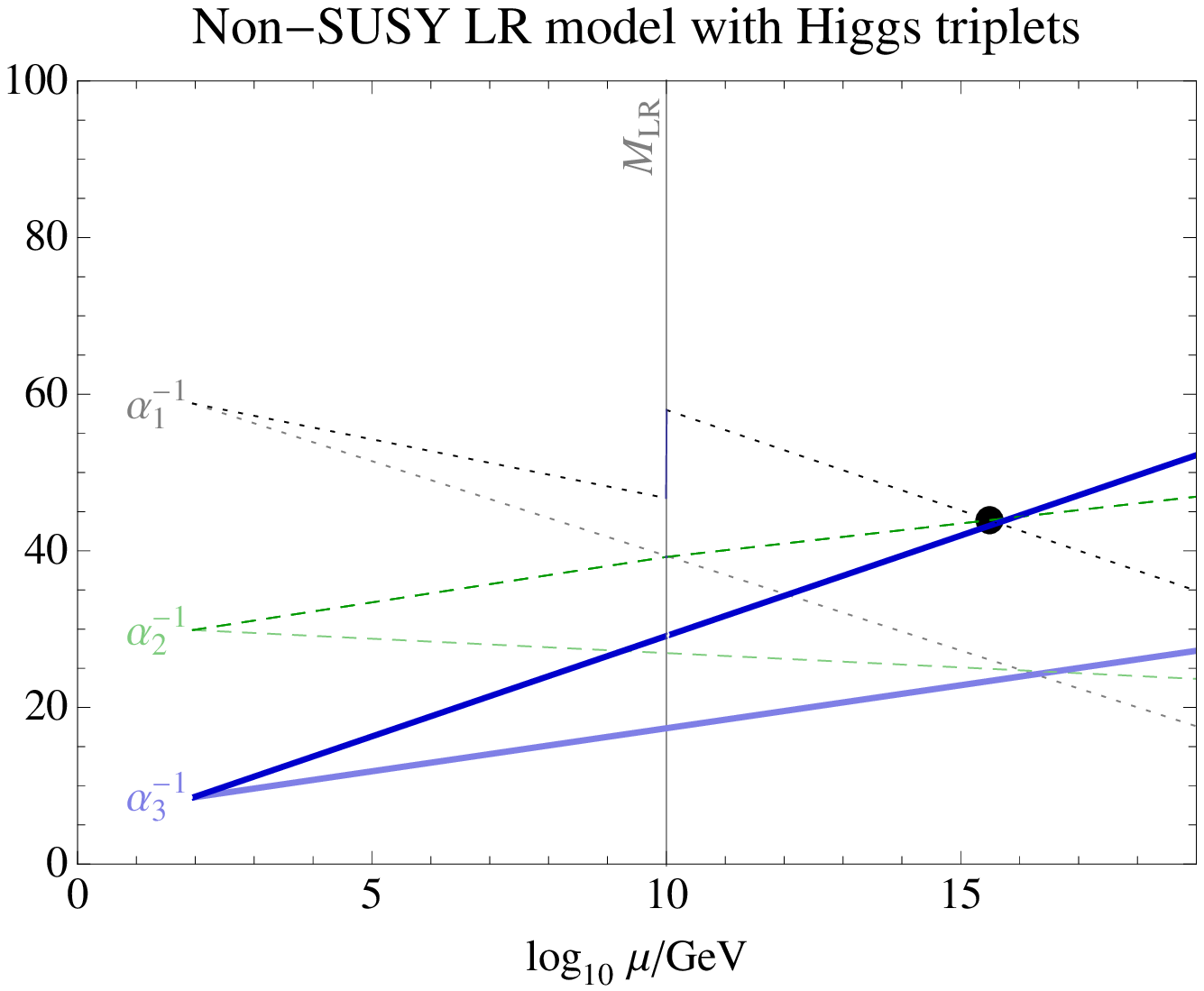} &
      \includegraphics[width=8cm]{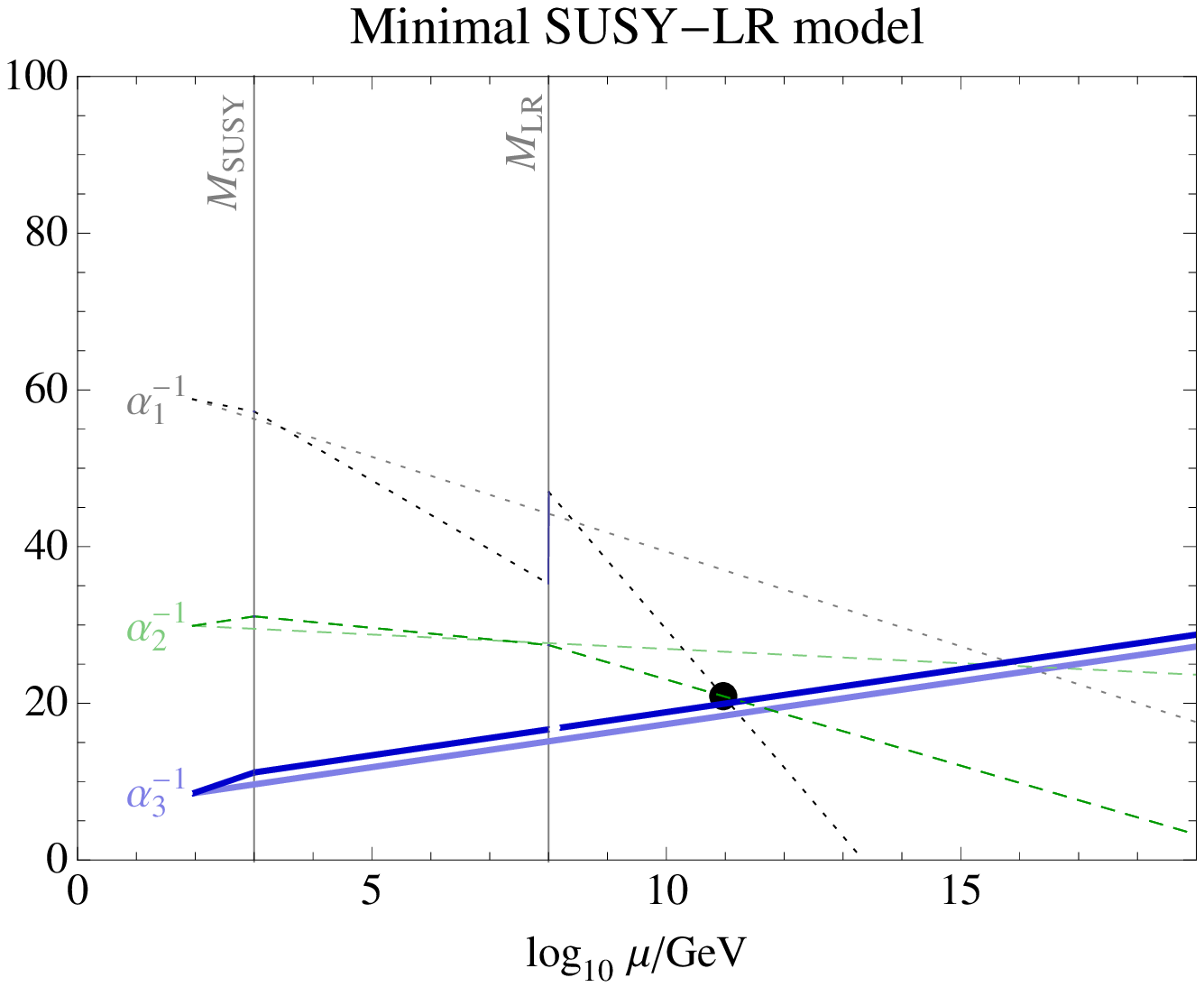}\\
     (i) & (ii)
    \end{tabular}\\[0.5cm]
    \begin{tabular}{c}
      \includegraphics[width=8cm]{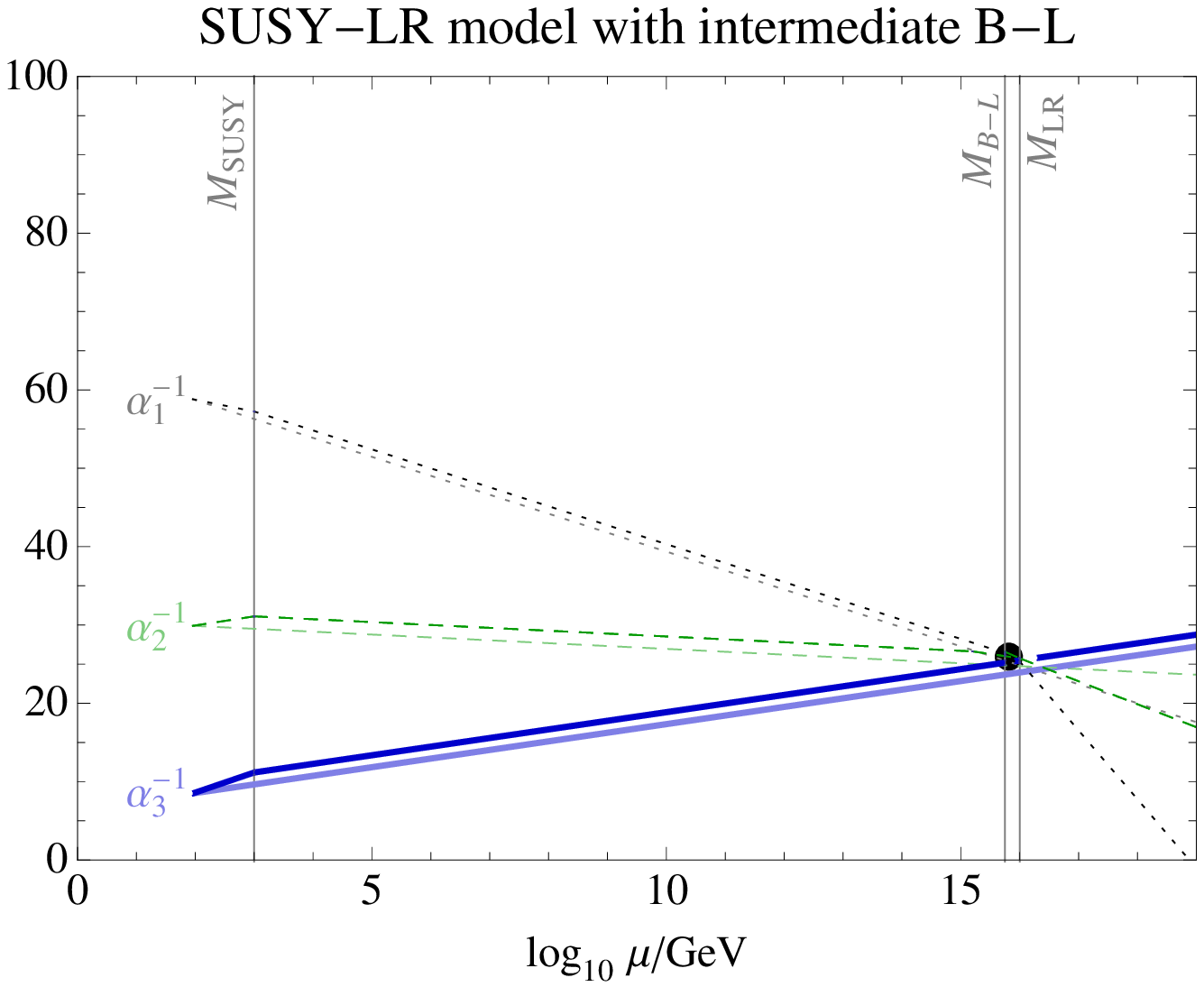} \\
      (iii)\\
    \end{tabular}
  \end{center}
  \caption{Renormalization group evolution in (i) a non-supersymmetric
    left-right model~\cite{Deshpande:1990ip}, (ii) the minimal supersymmetric LR
    model~\cite{Babu:2008ep}, and (iii) a non-minimal SUSY-LR
    model~\cite{Aulakh:1997ba,Aulakh:1997fq}. The light curves in the background
    correspond to the renormalization group running in the MSSM.}
  \label{fig:rge-lr}
\end{figure}

In fig.~\ref{fig:rge-lr} we compare the one-loop renormalization group running
of the three considered models and of the MSSM.  The embedding of the SM or
MSSM into the standard GUT groups $SU(5)$ and $SO(10)$ requires
$\alpha_3(M_{\rm GUT}) = \alpha_2(M_{\rm GUT}) = \frac{20}{3} \,
\alpha_1(M_{\rm GUT})$, but for the graphical presentation we have absorbed
the factor $\frac{20}{3}$ into the definition of $\alpha_1$, so that at
$M_{\rm GUT}$ the curves for $\alpha_1$, $\alpha_2$, and $\alpha_3$
meet at one point.  In the left-right models, the GUT normalization factor is
$\frac{8}{3}$ instead of $\frac{20}{3}$.  The matching condition for the
GUT-normalized $U(1)$ coupling constants at $M_{\rm LR}$ reads for models (i)
and (ii)
\begin{align}
  \alpha_{1, \rm LR}(M_{\rm LR}) = \frac{2}{5}
    \frac{\alpha_{1, \rm SM}(M_{\rm LR}) \, \alpha_2(M_{\rm LR})}
         {\alpha_2(M_{\rm LR}) - \frac{3}{5} \alpha_{1, \rm SM}(M_{\rm LR})} \,,
  \label{eq:matching}
\end{align}
where $\alpha_{1, \rm LR}$ is the $U(1)_{B-L}$ coupling constant, while
$\alpha_{1, \rm SM}$ and $\alpha_2$ correspond to $U(1)_Y$ and
$SU(2)_L$, respectively. In model (iii), a condition of the form
\eqref{eq:matching} is imposed not at $M_{\rm LR}$ but at $M_{B-L}$, with
$\alpha_2$ replaced by the $U(1)_R$ coupling constant.

Fig.~\ref{fig:rge-lr} shows that in all three models, unification is possible,
but, especially in case (ii), tends to occur at rather low scales, in possible
conflict with bounds from proton decay.  In fact, from dimensional analysis, we
expect the proton lifetime to be
\begin{align}
  \tau_p \sim \frac{M_{\rm GUT}^4}{m_p^5} \,.
  \label{eq:tau-p}
\end{align}
The bound $\tau_p > 2.1 \cdot 10^{29}$~yrs~\cite{Amsler:2008zz} then implies
$M_{\rm GUT} \geq \Mpd \equiv 10^{16}$~GeV. Therefore, models with $M_{\rm GUT}
< \Mpd$ can be embedded into a Grand Unified Theory only if special measures
are taken to forbid or suppress proton decay operators beyond the estimate
\eqref{eq:tau-p}.  Note also that the unified coupling constant $\alpha_{\rm
GUT}$ has a much larger value in the supersymmetric models than in the non-SUSY
case. The reason is that  according to eqs.~\eqref{eq:bi} and
\eqref{eq:bi-SUSY} the additional particle content of SUSY models always
\emph{increases} the beta function coefficients $b_i$.

In general, in supersymmetric models, besides the (dimension-six) operators
induced by X and Y gauge boson exchange, proton decay can be induced by
additional dimension-five operators arising from the exchange of coloured
higgsinos.  These are really dangerous operators, since they lead to extremely
fast proton decay. Indeed, if they are present, the proton life-time $\tau_p$
becomes proportional only to the second power in the GUT scale, instead of the
fourth power, as reported in eq.~$\eqref{eq:tau-p}$.

The purely supersymmetric contributions to proton decay have already been used
to set limits on SUSY-GUT models. For example, the minimal supersymmetric
$SU(5)$ GUT model has been tightly constrained by the Super-Kamiokande lower
bound on $p\rightarrow K^{+}\bar{\nu}$ decay channel~\cite{Murayama:2001ur},
assuming that the gauge coupling unification is satisfied. Several works have
dealt with the possibility of suppressing dimension-five operators. Some of
these models invoke extra-dimensions, see
e.g.~\cite{Nomura:2001mf,Dermisek:2001hp} or a more complicated higgs
sector~\cite{Babu:1993we,Babu:2002fsa}.

However, in our work we decided to pursue a conservative approach and thus
apply only the constraint on the unification scale derived through
eq.~$\eqref{eq:tau-p}$. It is, however, possible that some models with a
unification scale $M_{\rm GUT} \gtrsim \Mpd$ could be excluded by rapid proton
decay induced by dimension-five operators
\footnote{In principle, beyond gauge boson and higgsino exchange, two other
  sources of proton decay can be present: R-parity violating terms and
  dimension-five Planck suppressed operators.  However, these operators are
  not directly related to the unification scale (since they can also be
  present without unification), and therefore they do not provide a
  model-independent constraint on the value of $M_{\rm GUT}$.}.

\begin{figure}
  \begin{center}
    \includegraphics[width=0.45\textwidth]{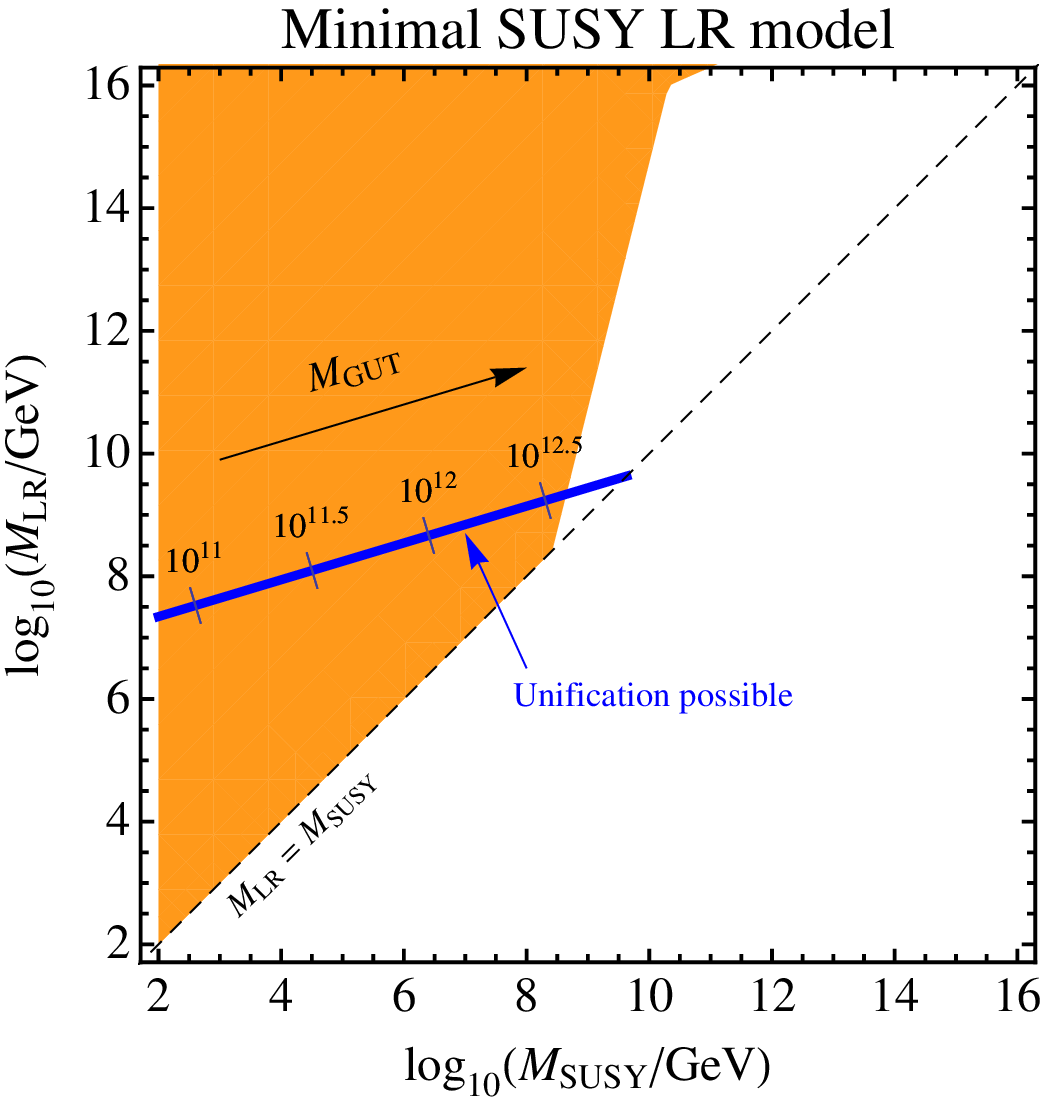}\hspace{5mm}\includegraphics[width=0.45\textwidth]{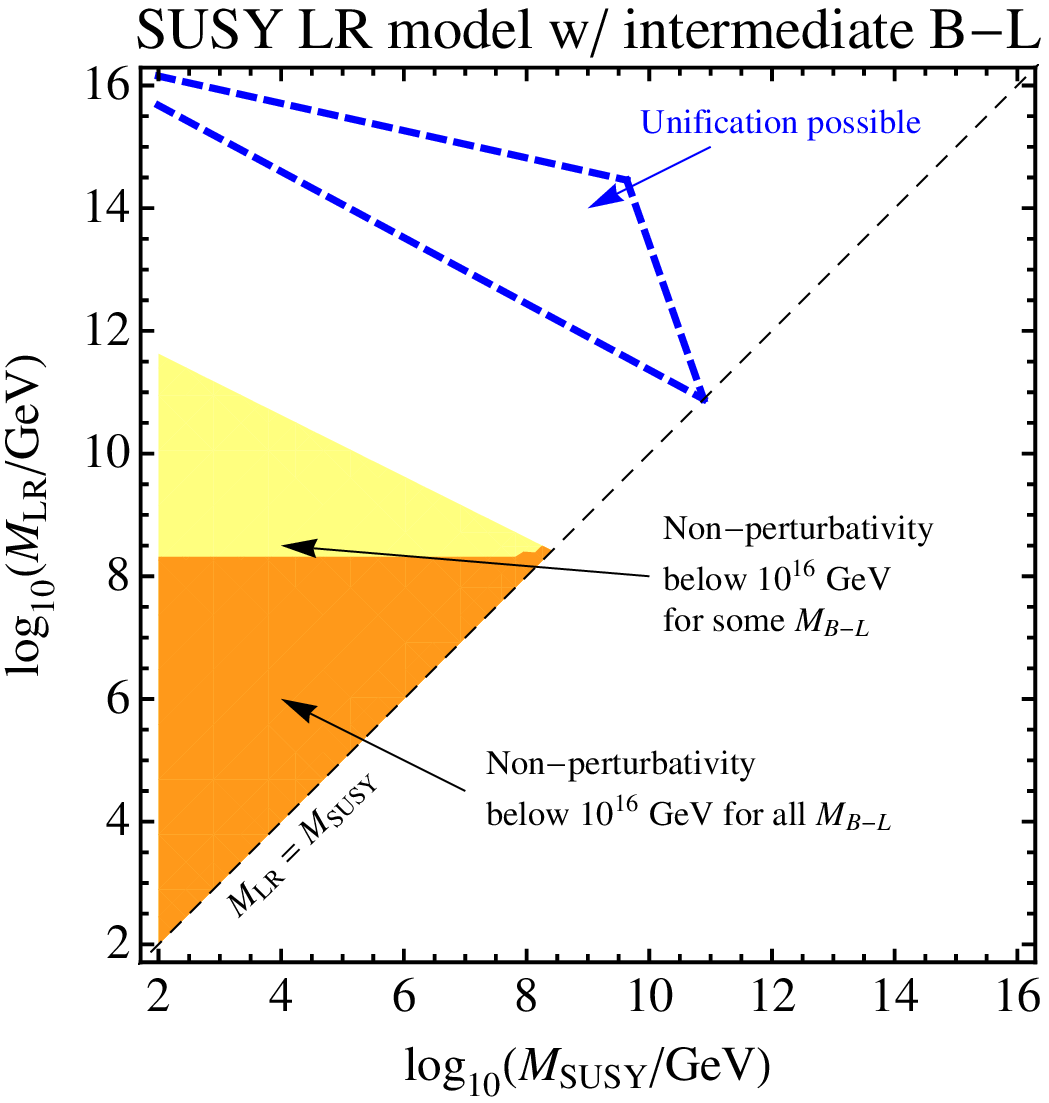}
    \\[5mm]
    \includegraphics[width=0.45\textwidth]{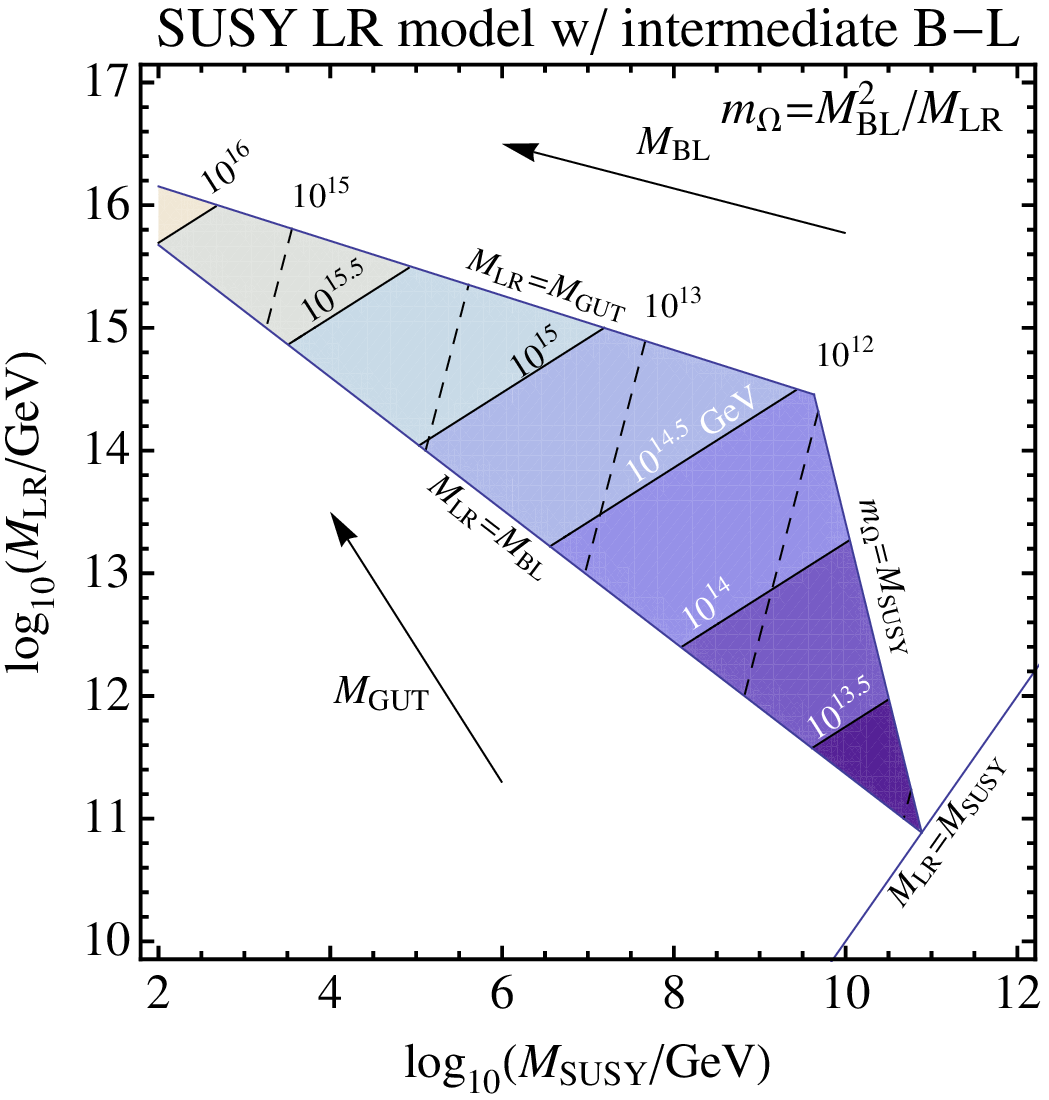}\hspace{5mm}\includegraphics[width=0.45\textwidth]{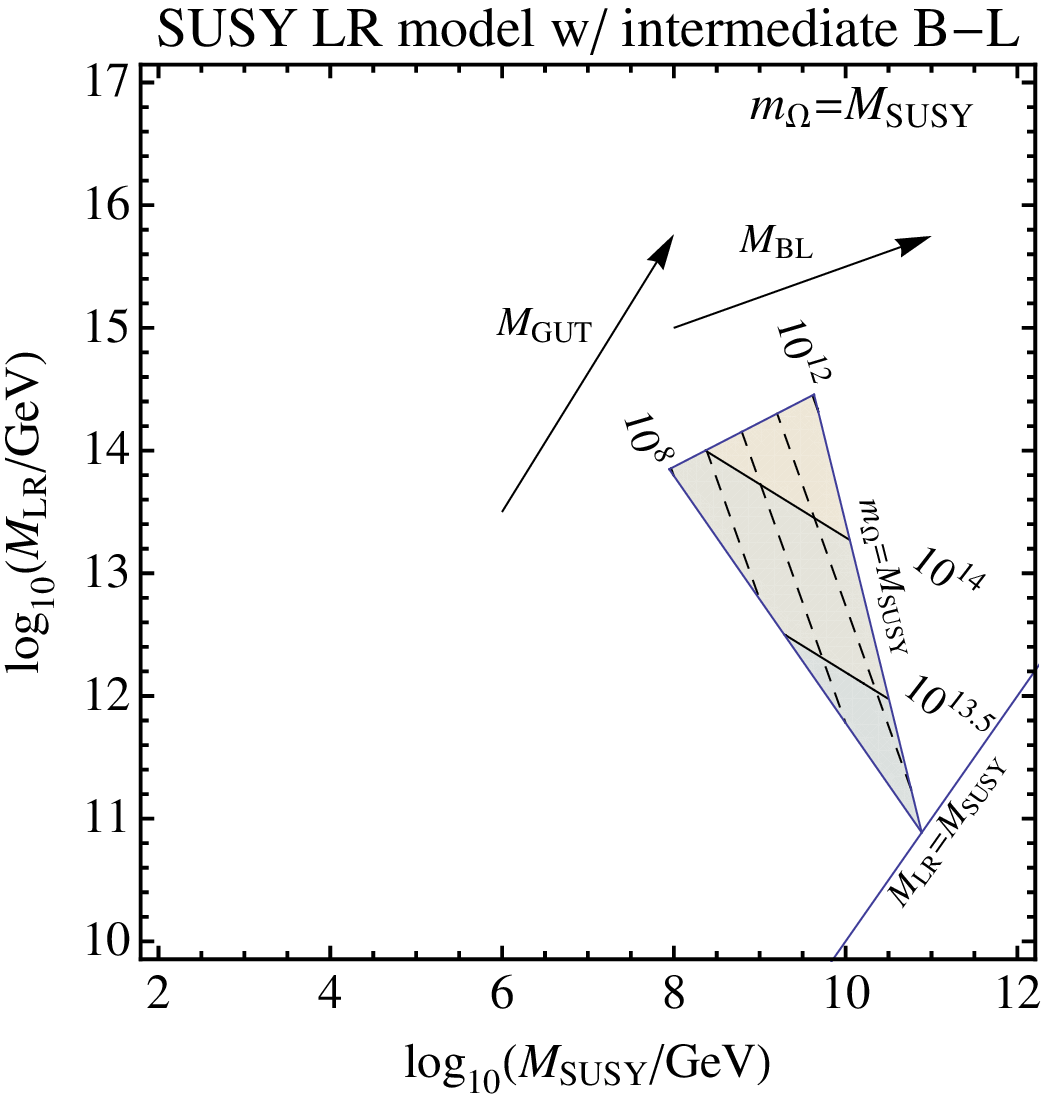}
  \end{center}
  \caption{Gauge coupling unification and non-perturbativity constraints on
    left-right symmetric models. The regions of parameter space leading to
    successful unification are marked on each plot, along with the unification
    scale, $M_{\rm GUT}$. The dark orange areas depict combinations of $M_{\rm
      SUSY}$ and $M_{\rm LR}$ for which the model becomes non-perturbative
    below the proton decay scale, $\Mpd$, so that even increasing the particle
    content at some intermediate scale will not be able to push $M_{\rm GUT}$
    above $\Mpd$ without violating perturbativity. For the model with
    intermediate $B-L$ breaking we also show scenarios where
    non-perturbativity below $\Mpd$~GeV occurs only for some choices of
    $M_{B-L}$ (light yellow area). In addition, the lower panels show the GUT
    scale (coloured shaded contours) and the $B-L$ breaking scale (dashed
    contours) required to achieve unification in this model. The lower left
    panel applies to the model where the mass of the SU(2)$_{L}$ triplet,
    $m_{\Omega} = M_{B-L}^2/M_{LR}$ and the right panel applies to the case
    where $m_{\Omega} = M_{\rm SUSY}$.}
  \label{fig:lr-scan}
\end{figure}

Let us now examine how varying the scales $M_{\rm LR}$, $M_{\rm SUSY}$, and
$M_{B-L}$ affects the prospects of Grand Unification in left-right-symmetric
models. For the non-SUSY model (i), only the choice $M_{\rm LR} \sim
10^{10}$~GeV (shown in fig.~\ref{fig:rge-lr}) leads to unification. For the
SUSY models, the indicated areas in fig.~\ref{fig:lr-scan} show for which
combinations of $M_{\rm LR}$ and $M_{\rm SUSY}$ unification occurs.  (For
model (iii) (right panel), for given values of $M_{\rm LR}$ and $M_{\rm
  SUSY}$, we chose the unique value $M_{B-L}$ in the range $[M_{\rm SUSY},
  M_{\rm LR}]$ which leads to unification). The GUT scale is marked on each
plot either explicitly along the line of values leading to unification (model
(ii)), or through the shaded contours (model (iii)). Notice that in most of
the parameter space unification is only possible in a narrow band of $M_{\rm
  SUSY}$ and $M_{\rm LR}$ values, and that in virtually all of these cases we
find $M_{\rm GUT} < \Mpd$, thus causing potential problems with proton decay.

Let us also remark that we have not found any solutions with $M_{\rm GUT} \sim
M_{\rm Pl}$, which would have been an interesting feature in the context of
quantum gravity theories. For model (iii), all unifying solutions we have found
correspond to $M_{LR} \sim M_{\rm GUT}$. (This observation is similar to what
has been found in~\cite{Bajc:2004xe} for a class of $SO(10)$ GUTs.)

One might hope to reconcile LR symmetry with Grand Unification above $\Mpd$
(or even at $M_{\rm Pl}$) by extending the particle content of the LR model.
In particular, the addition of extra coloured particles could postpone
unification, but, as discussed above, any new particle will inevitably bring
the model closer to non-perturbativity. The orange shaded regions in the upper
panels of fig.~\ref{fig:lr-scan} show for which values of $M_{\rm SUSY}$ and
$M_{\rm LR}$ it is definitely impossible to reconcile unification,
perturbativity and the proton decay bounds by adding extra matter because at
least one of the $\alpha_i$ becomes non-perturbative below $\Mpd$, even
without additional particles in the model. For model (iii) there is also a
yellow region in which this happens only for some choices of $M_{B-L}$. In
these regions of parameter space, any attempt to increase $M_{\rm GUT}$ by
adding new scalar or fermionic particles would be in even greater conflict
with perturbativity. We see that the problem is particularly severe in the
minimal SUSY-LR model (ii).  The reason is that this model has many particles
with low masses around $M_{\rm SUSY}$. In particular, the doubly charged
scalars $\delta^{c--}$ and $\bar{\delta}^{c++}$ have a very strong impact on
the running of $\alpha_1$.

\section{Grand Unification in a SUSY Pati-Salam model}
\label{sec:PS}

Let us now investigate Grand Unification in another well-motivated class of
models, namely those of the Pati-Salam (PS) type~\cite{Pati:1974yy} with the
gauge group $SU(2)_L \times SU(2)_R \times SU(4)$. In particular, we will
study the minimal supersymmetric PS model discussed in
\cite{Melfo:2003xi}. The gauge symmetry breaking pattern of this model is
\begin{align}
  & SU(2)_L \times SU(2)_R \times SU(4)
                                        \nonumber\\
  &\quad
  \xrightarrow{M_{\rm PS}}
  SU(3)_c \times SU(2)_L \times SU(2)_R \times U(1)_{B-L}
                                        \nonumber\\
  &\quad
  \xrightarrow{M_{\rm LR}}
  SU(3)_c \times SU(2)_L \times U(1)_Y \,,
  \label{eq:PS-breaking}
\end{align}
and SUSY is broken at $M_{\rm SUSY} < M_{\rm LR}$. The matter particles
reside in the representations
\begin{align}
  \psi(2, 1, 4)  \qquad\text{and}\qquad \psi^c(1, 2, 4^*)
  \label{eq:PS-matter}
\end{align}
of the PS gauge group. They get masses from the vevs of the Higgs bidoublets
\begin{align}
  \Phi(2,2,1) \qquad\text{and}\qquad \Phi(2,2,15)\,.
  \label{eq:PS-bidoublet}
\end{align}
Symmetry breaking is achieved by introducing Higgs multiplets
\begin{align}
  A(1, 1, 15)
  \label{eq:PS-A}
\end{align}
and
\begin{align}
  \Sigma(3, 1, 10)         \,,\quad
  \bar{\Sigma}(3, 1, 10^*) \,,\quad
  \Sigma^c(1, 3, 10^*)     \,,\quad
  \bar{\Sigma^c}(1, 3, 10) \,.
  \label{eq:PS-Sigma}
\end{align}
The particles surviving below $M_{\rm PS}$ are the usual matter particles, a
colour octet with mass $M_{\rm LR}^2 / M_{\rm PS}$ or $M_{\rm SUSY}$ (whichever
is larger) emerging from $A$, and two $SU(2)_R$ triplets $\Delta^c(1, 1, 3, -2)$
and $\bar{\Delta}^c(1, 1, 3, 2)$ of $SU(3)_c \times SU(2)_L \times SU(2)_R
\times U(1)_{B-L}$. The doubly charged components of $\Delta^c$ and
$\bar{\Delta}^c$ and a linear combination of their neutral components have
masses of order $M_{\rm SUSY}$, while the remaining components have masses of
order $M_{\rm LR}$.

\begin{figure}
  \begin{center}
      \includegraphics[width=8cm]{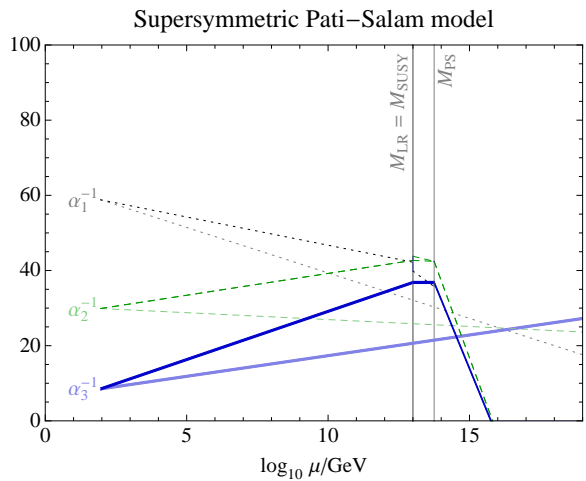}
  \end{center}
  \caption{Renormalization group evolution in the minimal SUSY Pati-Salam
  model~\cite{Melfo:2003xi}.}
  \label{fig:rge-ps}
\end{figure}

\begin{figure}
\begin{center}
\includegraphics[width=7.6cm]{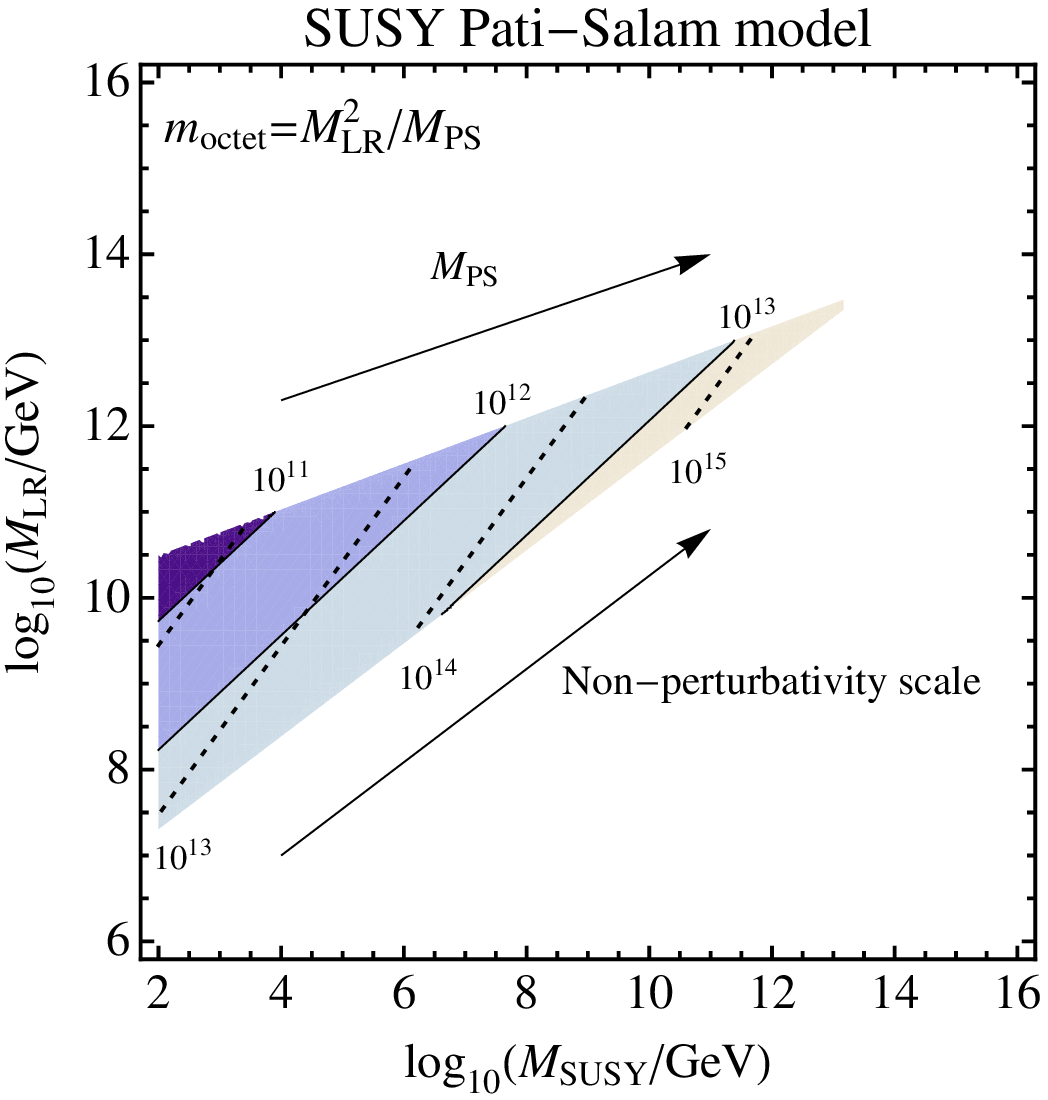}\hspace{5mm}
\includegraphics[width=7.6cm]{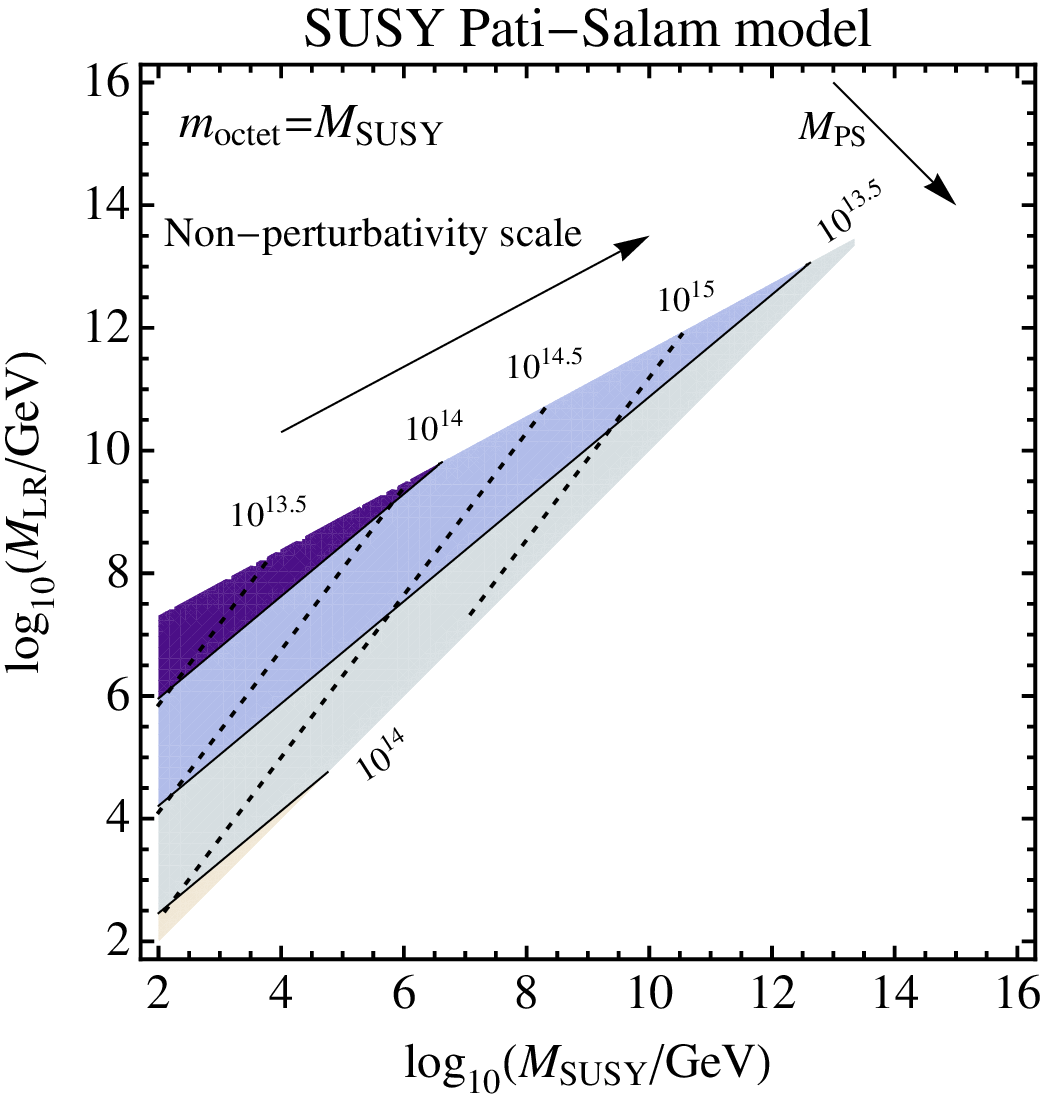}
\end{center}
\caption{Constraints on the minimal SUSY Pati-Salam model~\cite{Melfo:2003xi}
  coming from successful, perturbative, unification into the Pati-Salam
  group. The shaded region shows the part of parameter space where Pati-Salam
  unification is possible and the shaded contours indicate the Pati-Salam
  scale. The scale at which the model becomes non-perturbative is illustrated
  by the dashed contours.}
\label{fig:PSscales}
\end{figure}

An example for the running of the gauge couplings in the minimal SUSY PS model
is shown in fig.~\ref{fig:rge-ps}. After a detailed investigation of the
favoured values for $M_{\rm PS}$, $M_{\rm LR}$, and $M_{\rm SUSY}$ we find
that the Grand Unification of all gauge couplings is only possible in a very
narrow region of parameter space, and even then only by being flexible in the
uncertainty of $\alpha^{-1}_3 (M_Z)$. For the example in
fig.~\ref{fig:rge-ps}, Grand Unification is only possible for rather high
values of $M_{\rm SUSY}$, implying that SUSY cannot be considered as a
solution to the hierarchy problem here. Even partial unification into the
Pati-Salam group $SU(2)_L \times SU(2)_R \times SU(4)$ is only possible for
certain combinations of $M_{\rm SUSY}$ and $M_{LR}$, as shown in
fig.~\ref{fig:PSscales}. There, we also show the corresponding Pati-Salam
scales $M_{PS}$ (shaded contours), as well as the scales at which the
Pati-Salam coupling constants enter the non-perturbative regime (dotted
lines). We see that the non-perturbativity scale is always below $\Mpd$,
implying that the minimal SUSY Pati-Salam model cannot be further unified
without violating proton decay bounds (unless fundamentally new concepts such
as extra dimensions are introduced, see sec.~\ref{sec:ways-out}).

\section{Model-independent discussion of perturbativity and Grand Unification}
\label{sec:general}

Let us now generalize our findings from the previous sections to arbitrary
extensions of the Standard Model. The observation that models with large
particle content enter the non-perturbativity regime at relatively low scales
is quite generic (for exceptions, see sec.~\ref{sec:ways-out}) since it
follows directly from the fact that additional matter particles always
increase the coefficients $b_i$ (see equation~\eqref{eq:bi-SUSY}) Therefore,
if perturbativity up to the GUT scale is demanded in such models, gauge
coupling unification must also occur at relatively low scales, in tension with
proton decay bounds that suggest $M_{\rm GUT} \gtrsim \Mpd \sim 10^{16}$~GeV.
Thus, models with large particle content are disfavoured over more economic
ones.

\begin{figure}[p]
  \begin{center}
    \includegraphics[width=\textwidth]{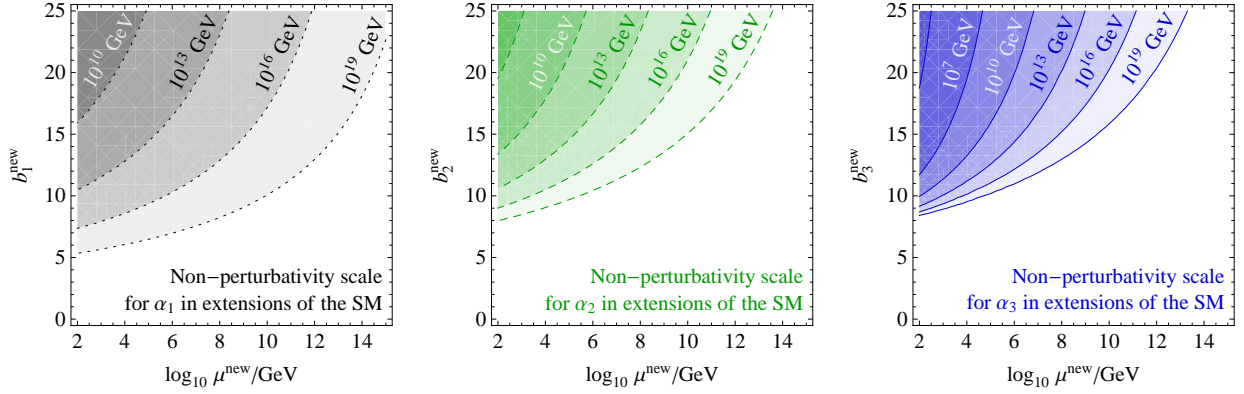}
  \end{center}
  \caption{Perturbativity constraints on extensions of the SM. We assume that
    new scalar or fermionic particles with masses of order $\mu^{\rm new}$
    are introduced and give contributions $b_i^{\rm new}$ to the $\beta$-function
    coefficients. The contours show the scale at which the one-loop approximations
    for the running coupling constants $\alpha_i$ become infinite
    as a function of $\mu^{\rm new}$ and $b_i^{\rm new}$.}
  \label{fig:NPScales-SM}
\end{figure}

\begin{figure}[p]
  \begin{center}
    \includegraphics[width=\textwidth]{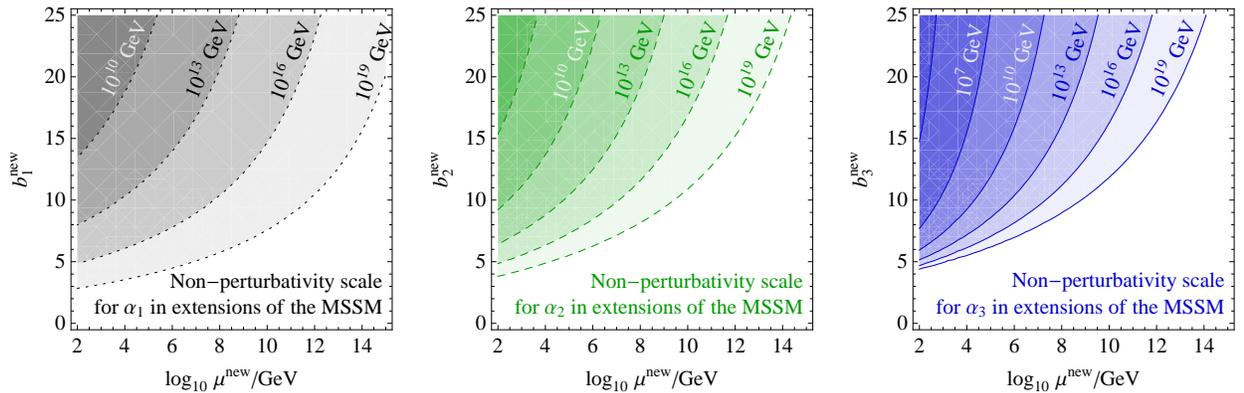}
  \end{center}
  \caption{Similar to fig.~\ref{fig:NPScales-SM}, but for extensions of the MSSM
    with $M_{\rm SUSY} \sim 10^2$~GeV.}
  \label{fig:NPScales-MSSM}
\end{figure}

To formulate the perturbativity constraints on models of new physics more
quantitatively, let us assume an arbitrary extensions of the SM or MSSM
particle content at a scale $\mu^{\rm new}$, with the new particles giving
contributions $b_i^{\rm new}$ to the $\beta$-function coefficients $b_i$.  The
three panels of fig.~\ref{fig:NPScales-SM} show the scales where $\alpha_1$,
$\alpha_2$, and $\alpha_3$ become infinite in the one-loop approximation as a
function of $\mu^{\rm new}$ and $b_i^{\rm new}$.  Fig.~\ref{fig:NPScales-MSSM}
shows similar results for the MSSM.  We see that for new physics at the TeV
scale, an increase of $b_1$ by 8 or of $b_2$ or $b_3$ by 9 would render the SM
non-perturbative below $\Mpd$, while for the MSSM this would happen already if
any of the $b_i^{\rm new}$ becomes larger than 5.  If new particles are
introduced well above the electroweak scale, the perturbativity constraints
become weaker.

In table~\ref{tab:beta}, we list the contributions to the $b_i^{\rm new}$ for
various hypothetical new particle representations.  We see that especially
large representations with high hypercharge are problematic: For example,
$\alpha_3$ would become non-perturbative below $10^{16}$~GeV if the SM is
extended by three vector-like colour octet quarks at 1~TeV.  One the other
hand, adding Higgs doublets or triplets to the SM is not a problem: $\alpha_2$
remains perturbative up to the Planck scale even if 48 new Higgs doublets or
12 Higgs triplets are added at the electroweak scale.  In the MSSM, there is
slightly less freedom to extend the particle content: Perturbativity of
$\alpha_2$ is lost below $10^{16}$~GeV if 10 additional $SU(2)$ doublets or 3
triplets are added, and $\alpha_1$ would run to $\infty$ below $10^{16}$~GeV
if 9 $Y=2$ superfields are introduced.

\begin{table}
  \centering
  \begin{tabular}{rccc}
    \toprule
    MSSM rep. & $b_1^{\rm new}$ & $b_2^{\rm new}$ & $b_3^{\rm new}$ \\
    \midrule
    $(Y, 1, 1)$ & $0.15 Y^2$ & $0$ & $0$ \\ 
$(Y, 2, 1)$ & $0.3 Y^2$ & $0.5$ & $0$ \\ 
$(Y, 3, 1)$ & $0.45 Y^2$ & $2$ & $0$ \\ 
$(Y, 4, 1)$ & $0.6 Y^2$ & $5$ & $0$ \\ 
$(Y, 5, 1)$ & $0.75 Y^2$ & $10$ & $0$ \\ 
$(Y, 6, 1)$ & $0.9 Y^2$ & $17.5$ & $0$ \\ 
$(Y, 7, 1)$ & $1.05 Y^2$ & $28$ & $0$ \\ 
$(Y, 1, 3)$ & $0.45 Y^2$ & $0$ & $0.5$ \\ 
$(Y, 2, 3)$ & $0.9 Y^2$ & $1.5$ & $1$ \\ 
$(Y, 3, 3)$ & $1.35 Y^2$ & $6$ & $1.5$ \\ 
$(Y, 4, 3)$ & $1.8 Y^2$ & $15$ & $2$ \\ 
$(Y, 5, 3)$ & $2.25 Y^2$ & $30$ & $2.5$ \\ 
$(Y, 6, 3)$ & $2.7 Y^2$ & $52.5$ & $3$ \\ 
$(Y, 7, 3)$ & $3.15 Y^2$ & $84$ & $3.5$ \\ 
$(Y, 1, 6)$ & $0.9 Y^2$ & $0$ & $2.5$ \\ 
$(Y, 2, 6)$ & $1.8 Y^2$ & $3$ & $5$ \\ 
$(Y, 3, 6)$ & $2.7 Y^2$ & $12$ & $7.5$ \\ 
$(Y, 4, 6)$ & $3.6 Y^2$ & $30$ & $10$ \\ 
$(Y, 5, 6)$ & $4.5 Y^2$ & $60$ & $12.5$ \\ 
$(Y, 6, 6)$ & $5.4 Y^2$ & $105$ & $15$ \\ 
$(Y, 7, 6)$ & $6.3 Y^2$ & $168$ & $17.5$ \\ 
$(Y, 1, 8)$ & $1.2 Y^2$ & $0$ & $3$ \\ 
$(Y, 2, 8)$ & $2.4 Y^2$ & $4$ & $6$ \\ 
$(Y, 3, 8)$ & $3.6 Y^2$ & $16$ & $9$ \\ 
$(Y, 4, 8)$ & $4.8 Y^2$ & $40$ & $12$ \\ 
$(Y, 5, 8)$ & $6 Y^2$ & $80$ & $15$ \\ 
$(Y, 6, 8)$ & $7.2 Y^2$ & $140$ & $18$ \\ 
$(Y, 7, 8)$ & $8.4 Y^2$ & $224$ & $21$ \\
    \bottomrule
  \end{tabular}
  \caption{Contributions of hypothetical new particles to the $\beta$-function
    coefficients $b_i$. Numbers are given for chiral superfields; in the
    non-SUSY case, they have to be multiplied by $1/3$ for complex scalars and by
    $2/3$ for chiral fermions. Note that the values for $b_1$ include the
    GUT normalization factor $\frac{3}{20}$.}
  \label{tab:beta}
\end{table}

\begin{figure}
  \begin{center}
    \includegraphics[width=\textwidth]{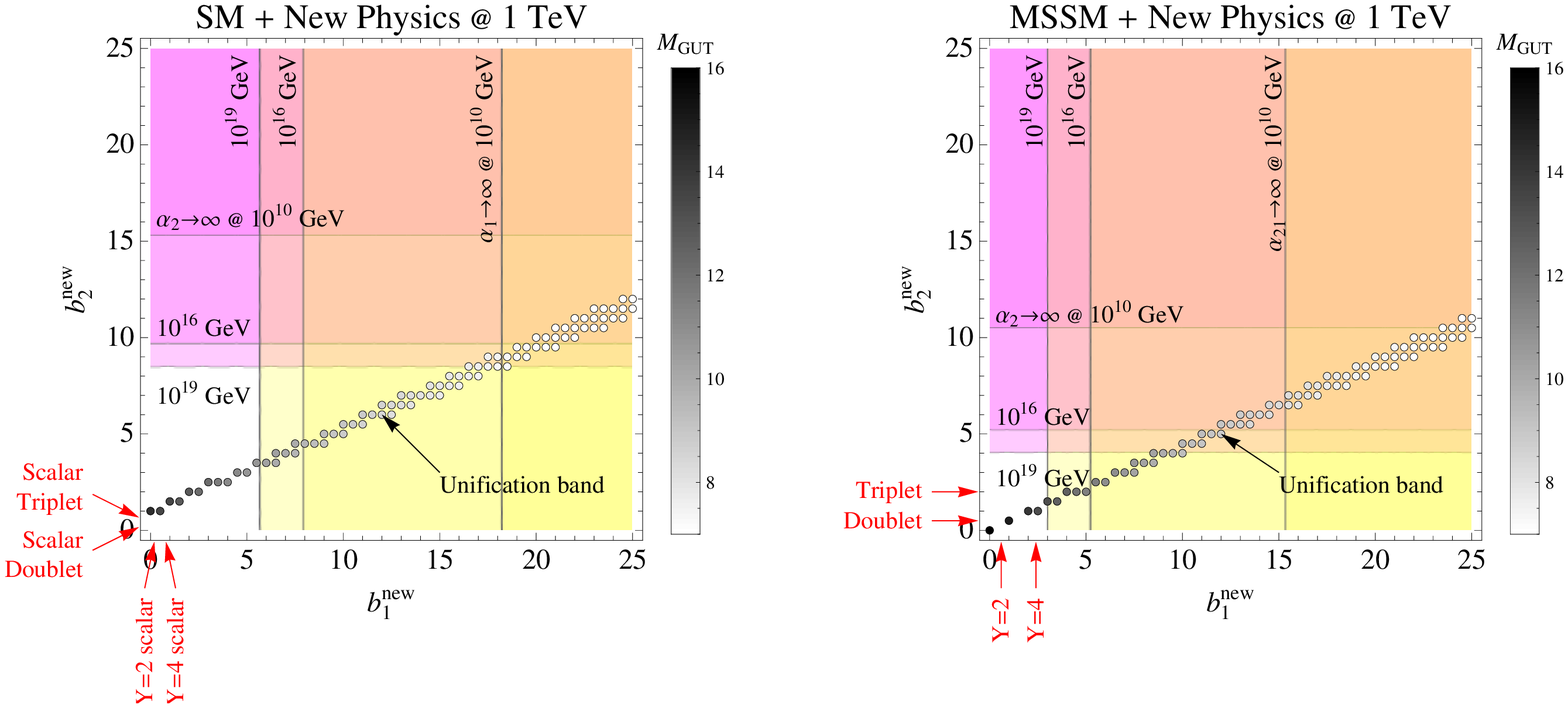}
  \end{center}
  \vspace{-1cm}
  \caption{Perturbativity and Grand Unification in extensions of the
    electroweak sector of the SM (left panel) and the MSSM (right panel),
    assuming new physics at 1~TeV. The shaded areas indicate the scale where
    $\alpha_1$ and $\alpha_2$ become non-perturbative as a function of the new
    particles' contributions to the beta function coefficients, $b_1^{\rm
      new}$ and $b^2_{\rm new}$.  The band of gray points shows for which
    combinations of $b_1^{\rm new}$ and $b_2^{\rm new}$ Grand Unification
    occurs. Dark points correspond to a high GUT scale, while lighter ones
    stand for low $M_{\rm GUT}$.}
  \label{fig:sm-scan}
\end{figure}

Many extensions of the SM focus especially on the electroweak sector, which
has much more room for interesting new phenomena at high energy than the QCD
sector.  Therefore, we will now consider models with no new coloured
particles, i.e.\ $b_3^{\rm new} = 0$, but with arbitrary contributions to
$b_1$ and $b_2$ from particles at the TeV scale. For models of this type, we
plot in fig.~\ref{fig:sm-scan} the non-perturbativity scales for $\alpha_1$
and $\alpha_2$ (shaded areas), and we indicate those combinations of $b_1^{\rm
  new}$ and $b_2^{\rm new}$ that lead to (perturbative) gauge coupling
unification (gray points). Dark points correspond to a high GUT scale, while
lighter ones stand for low $M_{\rm GUT}$. We see that Grand Unification is
possible in a certain band of $b_1^{\rm new}$ and $b_2^{\rm new}$ values. Of
course, even in those cases where no gauge coupling unification has been found
in our plot, it can be forced by adding suitably chosen coloured
representations to the model, provided that the points where $\alpha_1^{-1}$
and $\alpha_2^{-1}$ meet lies below the SM/MSSM curve for $\alpha_3^{-1}$, but
still in the perturbative regime.  We can also read from
fig.~\ref{fig:sm-scan} that for large beta function coefficients the GUT scale
is always rather low, in possible conflict with proton decay.  Again, one
might hope to increase $M_{\rm GUT}$ by adding more particles, especially
coloured ones, but this will inevitably lead to non-perturbative evolution if
the beta function coefficients become too large.

Thus we conclude that, for models in which the SM or the MSSM is extended
only by additional matter fields or Higgs particles at the electroweak scale,
Grand Unification above $\Mpd$ is only possible for $b_1^{\rm new}$
and $b_2^{\rm new}$ lying in the unshaded region of fig.~\ref{fig:sm-scan}.
Even then, in all cases except for the pure MSSM, pushing $M_{\rm GUT}$ above
$\Mpd$ requires exotic coloured particles.

\begin{figure}
  \begin{center}
    \includegraphics[width=\textwidth]{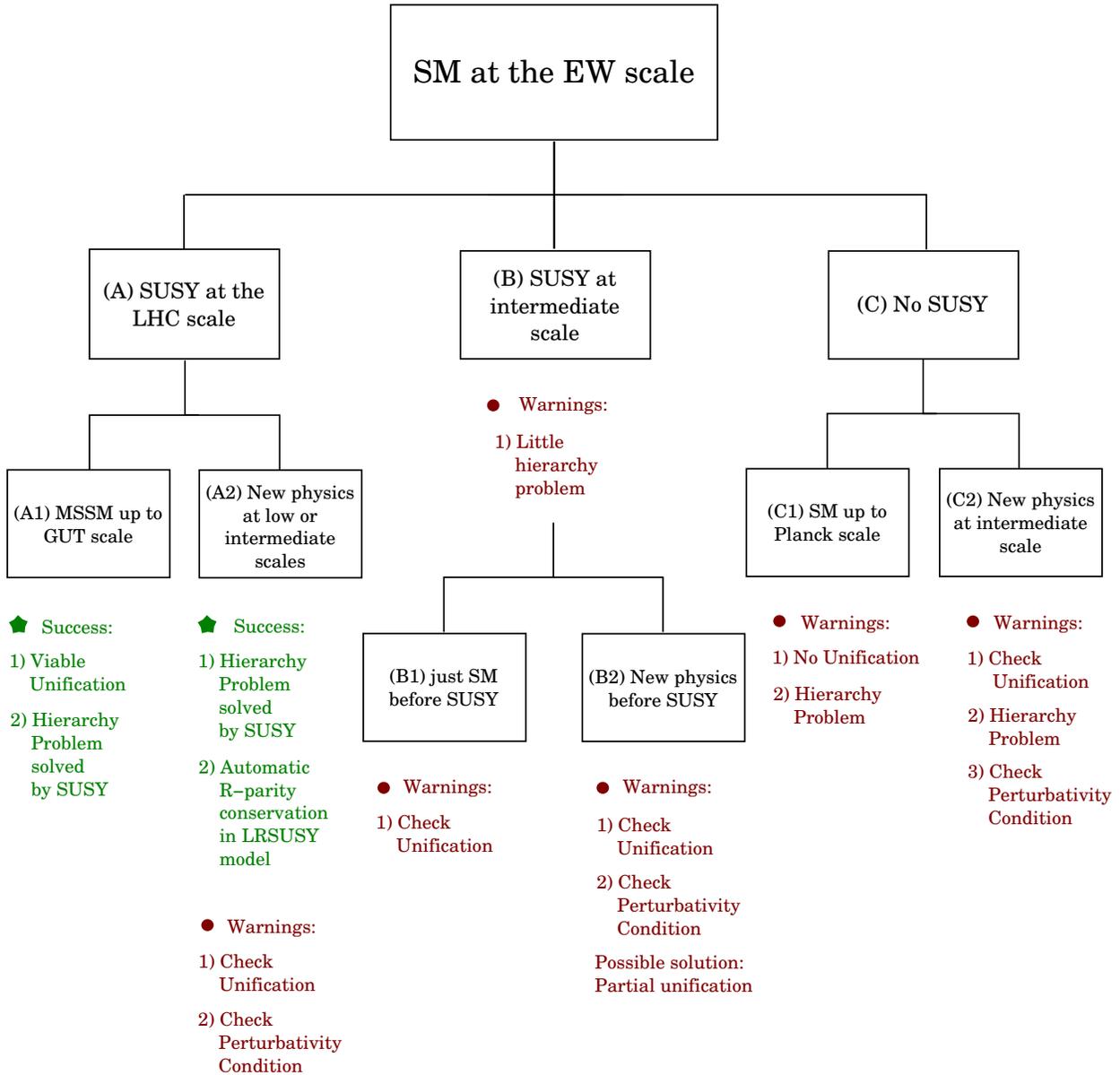}
  \end{center}
  \caption{A ``new physics'' flow chart.}
  \label{fig:chart}
\end{figure}

To end this section, we summarize the implications of perturbativity,
unification, and proton decay constraints for model building in the flow chart
shown in fig.~\ref{fig:chart}. Perturbativity is particularly problematic in
non-minimal SUSY models (scenarios (A2) and (B2)) because these tend to have
very large particle content. If SUSY does not exist up to $M_{\rm Pl}$ (cases
(C1) and (C2)), the perturbativity constraint can be fulfilled more easily,
but we encounter other difficulties, in particular the hierarchy
problem. Taking these considerations into account, the most attractive of the
considered scenarios is SUSY in its minimal form --- the MSSM (or,
equivalently, the NMSSM, which differs from the MSSM only by the addition of
one gauge singlet, see \cite{Maniatis:2009re} and references therein). Since
the MSSM does not provide gauge coupling unification and cannot solve the
hierarchy problem if $M_{\rm SUSY} \gg M_Z$ (case (B1)), we are left with case
(A1), the MSSM with $M_{\rm SUSY}$ around the LHC scale.

\section{Circumventing the perturbativity constraints}
\label{sec:ways-out}
In the analysis carried out in the previous sections, two main
problems with extensions of the Standard Model have appeared.

The first one is represented by the non-perturbative running of the gauge
couplings at high scales when the particle content is increased with respect
to the Standard Model or the MSSM. In figs.~\ref{fig:NPScales-SM}
and~\ref{fig:NPScales-MSSM} we have quantified the limits that the new
contributions to the $\beta$-function coefficients have to fulfil to preserve
perturbative coupling constants $\alpha_i$.

In principle, we could also accept the divergent evolution of the coupling
constants at high scales. However this means that we lose the ability to make
predications about physics above the non-perturbativity scale and, moreover,
we lose the ability to verify one of the main theoretical justifications for
physics beyond the SM, i.e. gauge coupling unification.

A possible way out could be provided by embedding the gauge group of the model
into a larger group at an intermediate scale $M_I$. This ``partial
unification'' will change the matter contribution $T_i(R)$ as well as the
gauge field contribution $C_{2i}$ in eqs.~\eqref{eq:bi} and \eqref{eq:bi-SUSY}
and could thus decrease $b_i$, keeping the model perturbative up to the Planck
scale. Moreover, the matching conditions at $M_I$ could be such that the
coupling of the new gauge group above $M_I$ is smaller than the corresponding
coupling constants of the SM, in the same way as the $\alpha_1$ corresponding
to $U(1)_{B-L}$ in the LR models can be smaller than the $\alpha_1$
corresponding to $U(1)_Y$ in the SM by virtue of eq.~\eqref{eq:matching}, see
fig.~\ref{fig:rge-lr} (i). We extensively investigate different classes of
left-right models in sec.~\ref{sec:LR} and Pati-Salam models in
sec.~\ref{sec:PS}. Our study shows that also in scenarios with partial
unification, one can easily run into similar perturbativity problems as in
models that preserve the SM gauge group up to high scales. Similar
difficulties are encountered in other models, for example the SUSY Little
Higgs model considered in ref.~\cite{Csaki:2005fc} which encounters
non-perturbativity below the unification scale when extra matter is added to
generate the top quark mass.

A realistic solution of the perturbativity problem is represented by
extra-dimension models in which only the gauge fields propagate in the bulk,
see e.g.~\cite{PerezLorenzana:1999qb}. In this case the number of gauge
degrees of freedom increases thus also increasing the negative contributions
to the $\beta$ function. The direct consequence is that the gauge couplings
are pulled away from the Landau pole and their perturbative evolution can be
followed up to the unification scale.

Low energy unification scales, $M_{\rm GUT}<\Mpd$ represent the second problem
that we experience in extensions of the Standard Model. Using the naive
dimensional estimate of eq.~\eqref{eq:tau-p}, it can be found that such low
values for $M_{\rm GUT}$ imply values of the proton life-time $\tau_p$ that do
not fulfil the experimental constraints provided by the Super-Kamiokande
detector~\cite{:2009gd,Kobayashi:2005pe}. However, the lower limit on the
unification scale can be relaxed if proton decay is forbidden or sufficiently
suppressed so that the proton lifetime $\tau_p$ becomes much larger than the
estimate of eq.~\eqref{eq:tau-p}.

One possibility to, in part, evade these experimental proton decay bounds is
to construct a ``contorted flavours'' GUT model \cite{Choi:2008zw}. Indeed,
there is no a-priori reason to couple the first quark family with the first
lepton family in a GUT multiplet. For example we could have the $(u,d)$ quarks
in the same representation as the $(\nu_{\tau},\tau)$ or $(\nu_{\mu},\mu)$. It
has been shown in~\cite{Choi:2008zw} that this pattern can lead to the correct
fermion masses and can kinematically suppress the proton decay channels into
charged leptons. However, the decay channels involving neutrinos are still
present and the experimental constraints reported
in~\cite{Kobayashi:2005pe,Amsler:2008zz} have to be taken into
consideration. In this case, it is not possible to completely remove the
constraints on the GUT scale arising from the experimental bounds on the
proton lifetime.

An economical way to avoid the constraint on the GUT scale is achieved by
extending the Higgs sector, see e.g.~\cite{PhysRevLett.43.893} for the case of
an SU(5) GUT. In this way the baryon number violating mixing matrix is, in
general, no longer related to the baryon conserving one and proton decay can
be suppressed by correctly adjusting the mixing angle in the baryon number
violating matrix. On the other hand, as we have shown, increasing the content
of the Higgs sector of a theory can easily spoil or make difficult the
perturbative unification of the gauge couplings, unless the the new Higgses
lie at the GUT scale.
 
Finally, we want to stress that the gauge coupling unification scale could be
different from the grand unification scale, as recently proposed
in~\cite{Kawamura:2009re}. In this model, the proton can naturally become
almost stable if the grand unification scale is big enough compared to the
scale where the gauge couplings meet.

\section{Conclusions}
\label{sec:conclusions}

In this paper, we have argued that the particle content of any extension of
the Standard Model or the MSSM is tightly constrained if gauge coupling
unification, perturbativity, and a GUT scale above the generic scale of proton
decay, $\Mpd \sim 10^{16}$~GeV, are demanded. For example, we have
demonstrated that in many left-right symmetric and Pati-Salam models, it is
impossible to fulfil all three requirements simultaneously
(cf.\ fig.\ \ref{fig:lr-scan}
: Either, one has to live with unification at lower scales and invent a
mechanism to circumvent proton decay bounds, or one has to extend the particle
content further in order to push $M_{\rm GUT}$ higher, but at the price of
losing perturbativity and thus predictivity. We have then generalized our
observations to a wider class of SM or MSSM extensions, and have examined the
constraints that perturbativity imposes on the scale of new physics and on its
contribution to the $\beta$-function coefficients
(cf.\ figs.~\ref{fig:NPScales-SM} and \ref{fig:NPScales-MSSM}).  Since extra
matter particles will always increase the values of the running gauge coupling
constants at high scales, our constraints favour extensions of the Standard
Model that require not too many new particles at low scales.  In other words,
the problems of the SM should be solved in a rather economic way.  In our
opinion, this is another hint in favour of the MSSM with $M_{\rm SUSY} \lesssim
1$~TeV.

We have finally discussed how the perturbativity constraints can be
circumvented either by simply accepting non-perturbativity, by designing models
with partial unification, by introducing extra dimensions, or by aiming for
Grand Unification at low to intermediate scales. In the latter case, special
measures have to be taken to forbid or suppress proton decay.

\section*{Acknowledgments}

We would like to thank J.\ W.\ F.\ Valle for very interesting and useful
discussions.  This work was in part supported by the Sonderforschungsbereich TR
27 `Neutrinos and Beyond' der Deutschen Forschungsgemeinschaft. JK would like
to acknowledge support from the Studienstiftung des Deutschen Volkes.

\begin{appendix}
\section{Analytic results for the minimal SUSY LR model}
In the following we define
\begin{equation}
t_a = \log\left(\frac{M_a}{M_Z}\right)\,,
\end{equation}
and firstly consider the minimal SUSY LR model detailed in
ref. \cite{Babu:2008ep}. After solving the 1-loop RGEs for the 4 gauge
couplings, we find that
\begin{equation}
  \alpha_3^{-1}(M_Z) = \frac{1}{4} \Big(9 \alpha_2^{-1}(M_Z)
- 5 \alpha_1^{-1}(M_Z) \Big) 
+ \frac{1}{8\pi} \Big(30 t_{LR} - 9 t_{\rm SUSY} \Big)\,.
\end{equation}
In this model the unification scale, $M_{\rm GUT}$, is given by
\begin{equation}
t_{\rm GUT } = \frac{2 \pi}{135} \Big(10 \alpha_1^{-1}(M_Z) - 3
  \alpha_2^{-1}(M_Z) - 7 \alpha_3^{-1}(M_Z)\Big) + \frac{71}{270} t_{\rm SUSY}\,.
\end{equation}

\section{Analytic results for the non-minimal LR Model with intermediate $B-L$
  scale}

We consider the non-minimal LR model of
refs. \cite{Aulakh:1997ba,Aulakh:1997fq} and assume the hierarchy $t_Z <
t_{\rm SUSY} < t_{BL} < t_{LR} < t_{\rm GUT}$, where $M_{\rm GUT}$ is the
unification scale. In addition we assume that $M^2_{B-L}/M_{LR} > M_{\rm
  SUSY}$, thus ensuring that the light SU(2)$_L$ triplet in this model has a
mass above the SUSY breaking scale (to be consistent with the results of
\cite{Aulakh:1997fq}).

After solving the 1-loop RGEs for the 4 gauge couplings, we find that
\begin{equation}
  \alpha_3^{-1}(M_Z) = \frac{1}{32} \Big(87 \alpha_2^{-1}(M_Z)
- 55 \alpha_1^{-1}(M_Z) \Big) 
+ \frac{1}{64\pi} \Big(216 t_{BL} - 36 t_{LR} + 97 t_{\rm SUSY} \Big)\,.
\end{equation}
Thus, given values of $M_{\rm SUSY}$ and $M_{LR}$, it is possible to calculate
the $M_{B-L}$ needed for successful unification, assuming the measured values
of the gauge couplings at $M_Z$. If this $M_{B-L}$ is self-consistent with the
assumptions above, then unification is possible for the specific choices of
$M_{\rm SUSY}$ and $M_{LR}$. In this case we find,
\begin{eqnarray}
t_{BL} & = & \frac{1}{216} \Big\{ 2\pi \Big( 55 \alpha_1^{-1}(M_Z) - 87
  \alpha_2^{-1}(M_Z) + 32 \alpha_3^{-1}(M_Z) \Big)  + 36 t_{LR} - 97
  t_{\rm SUSY}\Big\}\,,\\
t_{\rm GUT} & = & \frac{1}{864} \Big\{ 32\pi \Big( 5 \alpha_1^{-1}(M_Z) - 3
  \alpha_2^{-1}(M_Z) - 2 \alpha_3^{-1}(M_Z)\Big) + 288 t_{LR} - 128
  t_{\rm SUSY} \Big\}\,,
\end{eqnarray}
where the value of $\alpha_{\rm GUT}$ at the unification scale is
\begin{eqnarray}
\alpha_{\rm GUT}^{-1}(M_{\rm GUT}) & = & \frac{1}{72\pi} \Big\{ 4\pi \Big( 5 \alpha_1^{-1}(M_Z) - 3 \pi
  \alpha_2^{-1}(M_Z) + 16 \pi \alpha_3^{-1}(M_Z) \Big)\nonumber\\
& & \qquad \quad + 36 t_{LR} + 128 t_{\rm SUSY} \Big\}\,.\,\,
\end{eqnarray}

It is also possible to find solutions in this model with gauge coupling
unification but with $M_{B-L}$ such that $M^2_{B-L}/M_{LR} < M_{\rm SUSY}$. In
this case we assume that the light SU(2)$_L$ triplet acquires a mass at the
SUSY breaking scale $M_{\rm SUSY}$. This leads to different unification
conditions and the prediction for $\alpha_3^{-1} (M_Z)$ is now
\begin{equation}
\alpha^{-1}_3 (M_Z)  = \frac{1}{32} \Big( 87 \alpha_2^{-1}(M_Z) - 55
\alpha_1^{-1}(M_Z) \Big) + \frac{1}{64\pi} \Big( 138 t_{LR} - 132 t_{BL} + 271 t_{\rm
  SUSY} \Big)\,.
\end{equation}
As before, one can now predict $M_{B-L}$ and assuming that the condition
$t_{\rm SUSY}<t_{BL}<t_{LR}$ is met then successful unification is possible
with
\begin{eqnarray}
t_{BL} & = & \frac{1}{132} \Big\{ 2\pi \Big( 55 \alpha_1^{-1}(M_Z) -87
\alpha_2^{-1}(M_Z) + 32 \alpha_3^{-1}(M_Z)\Big) \nonumber\\
& & \qquad \quad + 138 t_{LR} + 271 t_{\rm
  SUSY} \Big) \Big\},\\
t_{\rm GUT} & = & \frac{1}{1056} \Big\{ 192\pi \Big(\alpha^{-1}_2(M_Z) - \alpha^{-1}_3(M_Z) \Big)
+ 480 t_{LR} + 208 t_{\rm SUSY} \Big \}\,,\\
\alpha^{-1}_{\rm GUT} (M_{\rm GUT}) & = & \frac{1}{88\pi} \Big\{ 8\pi \Big( 3 \alpha_2^{-1}(M_Z) + 8
\alpha_3^{-1}(M_Z) \Big) + 60 t_{LR} + 202 t_{\rm SUSY} \Big\}\,.
\end{eqnarray}

\section{Analytic results for the minimal Pati-Salam model}

In the minimal Pati-Salam model of ref. \cite{Melfo:2003xi}, we first assume
that $M^2_{LR}/M_{PS} > M_{\rm SUSY}$, thus ensuring the colour octet remains
heavier than $M_{\rm SUSY}$. After solving the RGEs, the following analytic
expressions are found
\begin{equation}
\alpha^{-1}_3 (M_Z) = \frac{1}{2} \Big( 5 \alpha^{-1}_1 (M_Z) - 3
\alpha^{-1}_2 (M_Z) \Big) + \frac{1}{\pi} \Big( 4 t_{\rm SUSY} - 12 t_{LR} - 9
t_{PS} \Big)\,,
\end{equation}
where $M_{PS}$ is the Pati-Salam symmetry breaking scale. Unification occurs
at
\begin{equation}
t_{\rm GUT} = \frac{5 \pi}{6} \Big( \alpha^{-1}_2 (M_Z) - \alpha^{-1}_1 (M_Z) \Big)+
t_{LR} + \frac{13}{3} t_{PS} - \frac{47}{36} t_{\rm SUSY}\,.
\end{equation}

In cases where  $M^2_{LR}/M_{PS} < M_{\rm SUSY}$, we instead assume that the
colour octet has a mass of $M_{\rm SUSY}$, and the following analytic
expression for $\alpha^{-1}_3 (M_Z)$ is found
\begin{equation}
\alpha^{-1}_3 (M_Z) = \frac{1}{2} \Big( 5 \alpha^{-1}_1 (M_Z) - 3
\alpha^{-1}_2 (M_Z) \Big) + \frac{1}{2 \pi} \Big( 5 t_{\rm SUSY} - 6 t_{LR} - 21
t_{PS} \Big)\,.
\end{equation}
The expression for the unification scale however remains unchanged.

\end{appendix}

\begin{center}
  \rule{10cm}{0.25pt}
\end{center}
\vspace{-1.5cm}
\bibliographystyle{apsrev}
\bibliography{./rge}

\end{document}